\documentclass[useAMS,usenatbib]{mn2e}
\usepackage{graphicx}

\title[Astronomical seeing and ground-layer turbulence in the Canadian High Arctic]
{Astronomical seeing and ground-layer turbulence in the Canadian High Arctic}

\author[P. Hickson, et al.]
{
P. Hickson$^{1,2,3}$\thanks{E-mail:hickson@physics.ubc.ca}, 
R. Gagn\'e$^{1}$, T. Pfrommer$^{2}$, E. Steinbring$^{4}$
\\
$^{1}$Department of Physics \& Astronomy, University of British Columbia, 
6224 Agricultural Road, Vancouver, B.C. V6T 1Z1, Canada\\
$^{2}$European Southern Observatory, Karl-Schwarzschild-Str. 2, Garching bei M\"unchen, Germany\\
$^{3}$Institut d'Astrophysique et de G\'eophysique, Universit\'e de Li\`ege, All\'ee du 6 ao\^ut, 
17 - B\^at. B5c, B-4000 Li\`ege 1, Belgique \\
$^{4}$National Research Council of Canada, 5071 W. Saanich Rd., Victoria, BC V9E 2E7, Canada
}

\begin{document}

\date{Accepted 2013 April 24. Received 2013 February 13; in original form 2013 February 13}

\pagerange{\pageref{firstpage}--\pageref{lastpage}} \pubyear{2012}

\maketitle

\label{firstpage}

\begin{abstract}
We report results of a two-year campaign of measurements, during arctic
winter darkness, of optical turbulence in the atmospheric boundary-layer 
above the Polar Environment Atmospheric Laboratory in northern Ellesmere 
Island (latitude +80$^\circ$ N). The data reveal that the ground-layer 
turbulence in the Arctic is often quite weak, even at the comparatively-low
610 m altitude of this site. The median and 25th percentile
ground-layer seeing, at a height of 20 m, are found to be 0.57  and 0.25 arcsec,
respectively. When combined with a free-atmosphere component of 0.30 arcsec,
the median and 25th percentile total seeing for this height is 0.68 and 
0.42 arcsec respectively. The median total seeing from a height of 
7 m is estimated to be 0.81 arcsec. These values are comparable to 
those found at the best high-altitude astronomical sites. 

\end{abstract}


\section{Introduction}

The high glacial plateau of Antarctica has attracted considerable interest from 
astronomers due to unique benefits offered by its geographic location. 
The long periods of continuous darkness available at extreme latitudes 
are important for time-sensitive programs such as exoplanet transit 
surveys and targets of opportunity. Low ambient temperatures result 
in reduced backgrounds for infrared observations.
In addition, there is evidence that the free 
atmosphere has relatively weak turbulence and potential for 
superlative astronomical seeing \citep{l2004,a2005,a2006}. 
The result is improved performance for telescopes for 
observations with and without adaptive optics.

The Arctic offers many of the same advantages as the Antarctic. But
it also has some additional attractions. Coastal mountain ranges
reach high elevations, and are
well away from central icecaps which potentially suffer the
strong surface-layer turbulence associated with the glacial plateau 
of Antarctica  \citep{m1996,m1999,m2002}. 
In addition, isolated unglaciated terrain can provide solid footings 
for astronomical telescopes. In northern Canada, some of these locations
are accessible from existing research bases.

The northernmost tip of Ellesmere Island is within $8^\circ$ of the
Pole. More than 50 years of continuous weather records from the
manned stations of Alert and Eureka show that the 
climate is dry, with typical winter temperatures averaging -40 C. In winter
darkness a strong inversion layer develops in the lower 1-2 km of the
atmosphere, producing stable atmospheric conditions. During that
time clear sky is found approximately 60\% of the time \citep{s2010}. 
Both Eureka and Alert have airstrips that are accessible year round 
by passenger and cargo aircraft, and Eureka is accessible by ship in 
the summer.

These considerations have motivated a program of in-situ measurements 
at several mountain sites with the aim of assessing the suitability of the region
for astronomy \citep{s2008,s2010}. Robotic instruments were deployed during
the summer by helicopter at three mountain sites near the north shore 
of Ellesmere Island. Their purpose was to measure weather and sky 
conditions through the winter. The data confirmed both the expected
high clear-sky frequency and low median wind speeds associated with
the development of a strong thermal inversion layer during winter darkness.

Most recently, our attention has focussed on
a mountain ridge near Eureka which is the site of a laboratory for climate 
research. The Polar Environment Atmospheric Research Laboratory (PEARL),
originally built for atmospheric ozone measurements, has been refurbished 
and used by the Canadian Network for the Detection of Atmospheric
Change since 2005. This facility has essential infrastructure, including electrical
power and a broadband satellite communications link. Until recently,
it was staffed by technicians year round. The combination of infrastructure,
accessibility, and a mountain-top location makes PEARL an ideal facility
for testing arctic instruments, and a potential site for astronomical telescopes.

The PEARL laboratory is located on a ridge, aligned north-south, at 610 m altitude
(see \citealt{s2010} for additional information and a map of the region).
The surrounding terrain falls steeply towards the south and more gradually 
towards the north and east. The principal topographic feature that could induce 
turbulence is a parallel ridge of equal altitude, approximately 200 m to the
west. This ridge can be clearly seen in the panoramic view
of Fig. 1. However, during good winter weather the wind rarely comes from 
this direction, the prevailing winds during clear skies are northerly, with 
bad weather usually arriving from the south.

\begin{figure*}
\includegraphics[width=175 mm]{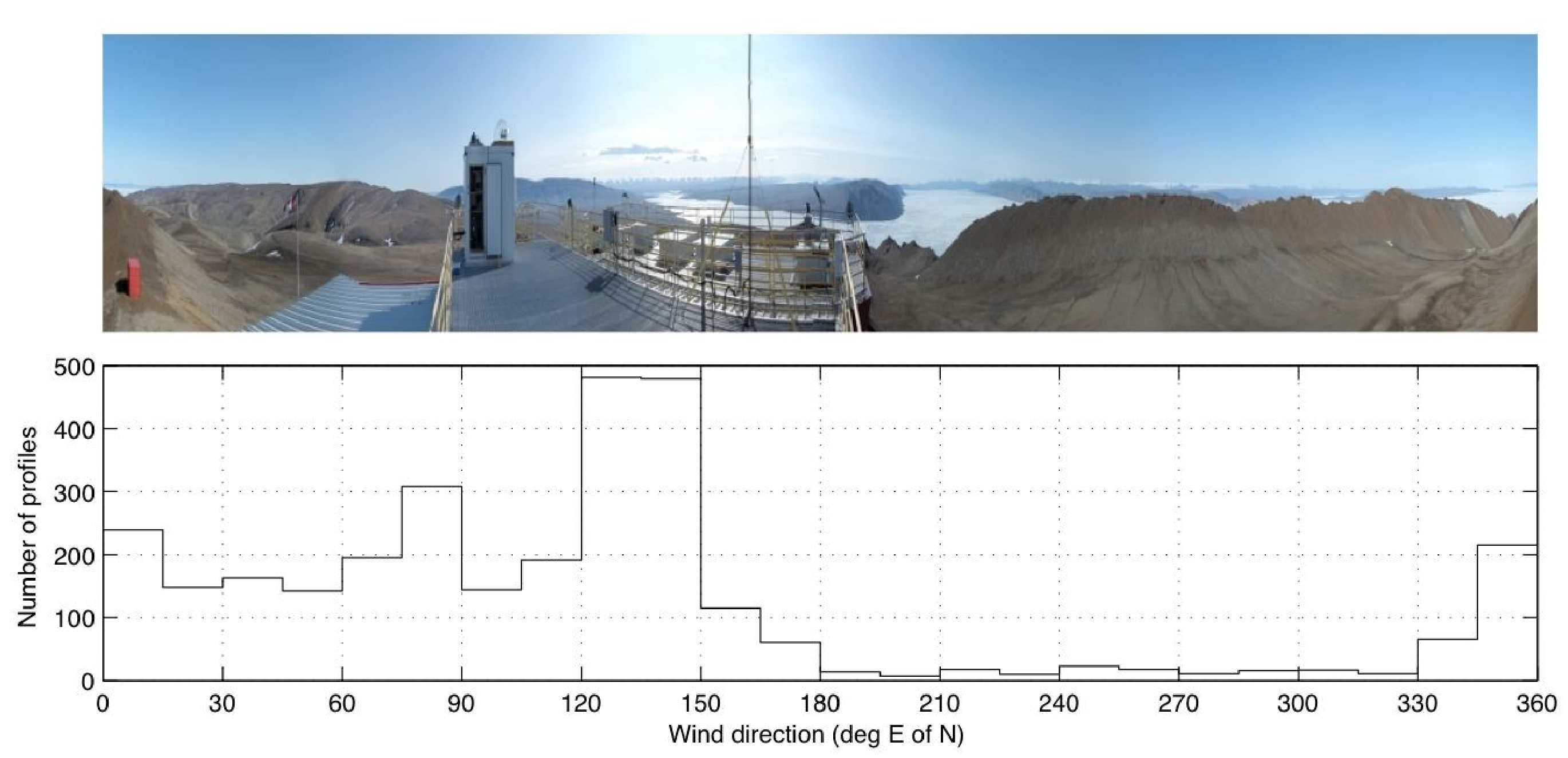}
%  \vspace{302pt}
 \caption{Panoramic view of the PEARL site, as seen during summer. 
 The top panel shows a 360-degree
 view as seen from the location of the ATP instrument on the NW corner
 of the roof of the PEARL laboratory building. The lower panel shows a
 histogram of wind direction during which which useful ATP data were 
 obtained.
 }
\end{figure*}

In 2009, an instrument was developed at the University of 
British Columbia for the express purpose of characterizing atmospheric
turbulence and astronomical seeing at remote Arctic sites - and designed
to operate under harsh environmental conditions. The
Arctic Turbulence Profiler (ATP) is a specialized lunar scintillometer 
which senses optical turbulence by measuring correlations in the scintillation 
of moonlight. From the data one obtains time-resolved profiles of turbulence 
strength as a function of altitude within the first km of the atmosphere. From this 
one can determine the contribution of ground-layer turbulence to the total seeing 
as a function of telescope height above ground.

As a first step towards eventual deployment at the remote northern mountain
sites, the ATP was operated at PEARL for two years, beginning in the fall of
2009. Initial results from the first winter were reported by \cite{h2010}. 
In this paper we present results from the full two-year data set. 

\section[]{Observations}

The ATP instrument has been described in detail by \cite{h2010}. 
It employs silicon photodiodes to measure the small fluctuations in 
the lunar flux that arise when the light propagates through atmospheric
turbulence. These fluctuations are spatially correlated, with a
coherence length that is proportional to the altitude of the turbulence.
By analyzing the coherence between pairs of photodiodes as a 
function of their mutual separation, it is possible to obtain an estimate
of the turbulence index of refraction structure constant, $C_N^2$,
as a function of height $h$ above the ground. A two-minute average 
is used to obtain accurate covariances, so the ATP produces 
a $C_N^2$ profile every 2 minutes.

The 48 photodiodes in the ATP are arranged in 6 rings
located at different heights along a vertical axis. Each ring contains
8 photodiodes separated in azimuth angle by 36$^\circ$
(a $72^\circ$ arc on the north side of the instrument is not filled because
the Moon is too low in the sky in that direction).  As the Earth rotates, 
the Moon illuminates sequentially one photodiode in each ring. A computer 
selects the appropriate photodiode for data collection. In this manner, the 
Moon can be tracked without any physical motion of the instrument. At any given time, six 
photodiodes (one per ring) record the irradiance (flux) of light from the Moon,
providing 15 independent pairs having separations (baselines) ranging from
0.128 to 2.000 m. The signal from each photodiode is amplified and both the 
steady (DC) and fluctuating (AC) signals are digitized. The sampling rate is
800 Hz. From this the dimensionless irradiance fluctuation $(I-<I>)/<I>$ is
determined for each of the photodiodes.
Data are recorded automatically whenever the Moon is more than 
19$^\circ$ above the horizon and when the Sun is more than 9$^\circ$ 
below the horizon, regardless of atmospheric conditions. 

Each photodiode is protected by an entrance window that has a conductive
coating. Power was periodically applied, to windows not currently observing 
the Moon, in order to melt any frost or ice. An optical filter, located between the 
window and the photodiode passes wavelengths greater than 700 nm. 
It's purpose is to block the strong auroral lines found at 558, 630, 636 and 656 nm.

The ATP was sent to PEARL in August 2009 and installed on the roof
of the laboratory. As Fig. 1 shows, this location has an unobstructed view of
the surrounding terrain. The bottom of the array was approximately 7 m above 
ground level. Engineering tests were conducted in September and 
October, and observations commenced on November 1. 
Fig. 2 shows the days and times of observations, and the corresponding lunar altitude. 
Observing statistics are reported in Table 1. Instrument problems interrupted 
the observations on three occasions (Nov 5 and 6, 2009 and 
Feb 19, 2010). The instrument was damaged during a storm in December 2009, 
which precluded any observations until after its repair in January 2010. 

\begin{figure}
\hspace{-6mm}
\includegraphics[width=98 mm]{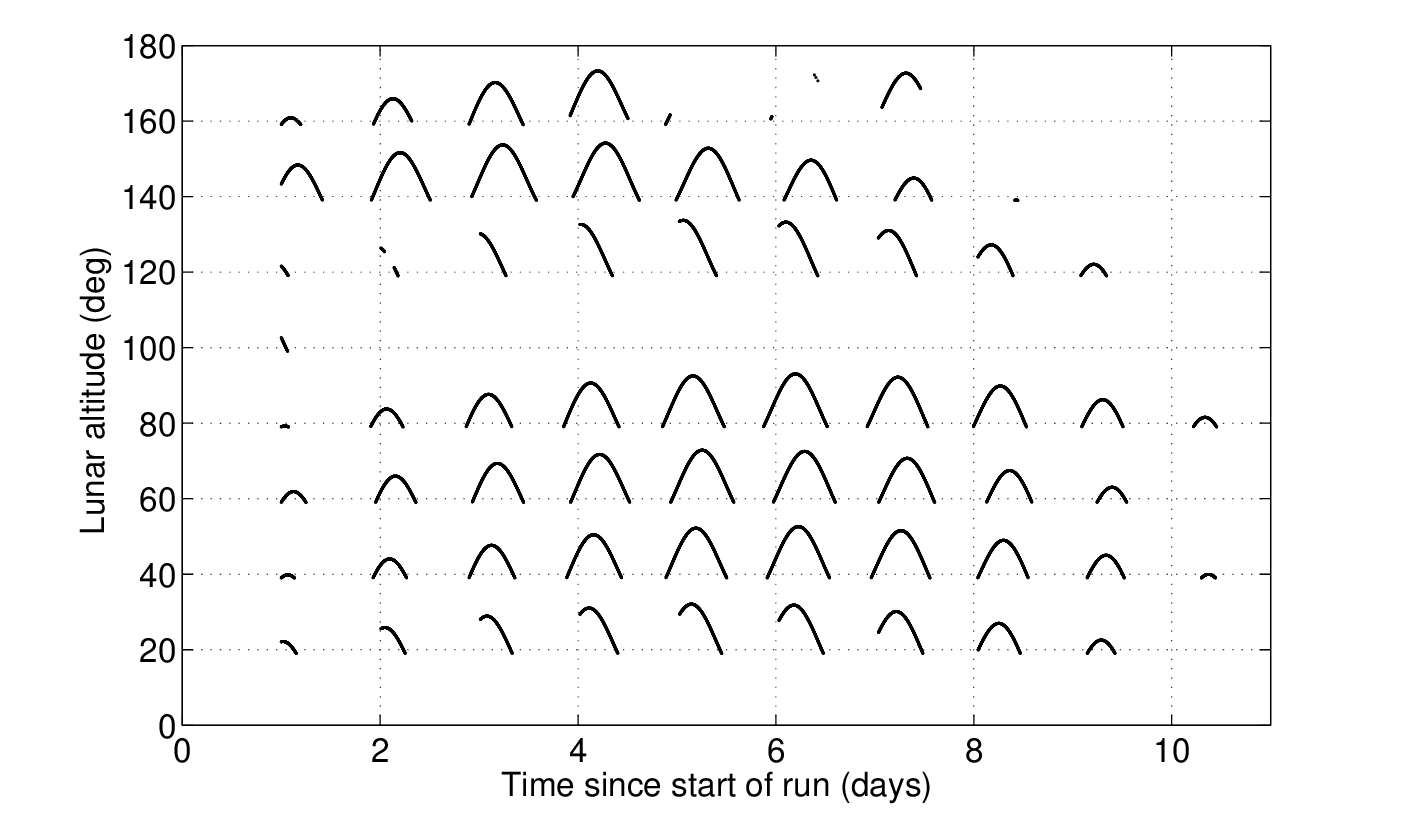}
%  \vspace{302pt}
 \caption{Lunar altitude is plotted vs time for all acquired data. Here time is
 measured in days, with 1 being the time of the first data sample of each run.
 Offsets of 20 degrees have been added to the vertical axis in order to separate 
 the individual runs. From top to bottom, the rows correspond to runs $1 - 8$ 
 in Table 1.}
\end{figure}

\begin{table*}
 \centering
 \begin{minipage}{110mm}
  \caption{Log of observations}
  \begin{tabular}{@{}llrrrr@{}}
  \hline
  (1) & (2) & (3) & (4) & (5) & (6) \\
  Run & Dates & Acquired & Frost-free & Clear & Photometric \\
  & & (hr) & (hr) & (\%) & (\%) \\
  \hline
  1 & 2009/11/01 - 2009/11/07 & 51.63 & 12.23 & 40.60 & 40.33 \\
  2 & 2010/01/23 - 2010/01/31 & 94.00 & 40.33 & 81.49 & 74.63 \\
  3 & 2010/02/18 - 2010/02/27 & 60.50 & 44.37 & 47.18 & 16.53 \\
  4 & 2010/03/22 - 2010/03/26 & 1.57 & 1.57 & 100.00 & 31.91 \\
  5 & 2010/11/18 - 2010/11/27 & 108.13 & 29.27 & 65.60 & 57.18 \\
  6 & 2010/12/15 - 2010/12/24 & 104.90 & 30.93 & 11.96 & 7.22 \\
  7 & 2011/01/11 - 2011/01 21 & 104.80 & 51.90 & 84.97 & 65.32 \\
  8 & 2011/11/01 - 2011/02/17 & 75.50 & 56.80 & 62.97 & 19.89 \\
  Total & 2009/11/01 - 2011/02/17 & 601.03 & 267.40 & 60.99 & 40.03 \\
\hline
\end{tabular}
\end{minipage}
\end{table*}

\section[]{Data analysis}

Although the ATP has heated windows, they were not always effective at removing 
frost and ice. The presence of frost can be detected by comparing the DC signals 
received from the 6 photodiodes that are illuminated by the Moon at any given time. 
If the windows are clear, these signals typically differ by less than 1\%. If frost or ice 
is present, these variations become much larger. A threshold of 3\% for the RMS 
variation in the DC signals was taken as being an indication of frost, ice or ice-fog. 
When possible, this was confirmed by direct observation by an onsite technician.

According to this ice criterion, 51\% of our data were found to be affected by 
frost, ice, or ice crystals and were therefore rejected. The number of frost-free 
hours for each run are reported in column (4) of Table 1.

For each 120-second data record, the expected lunar flux was estimated by means 
of a semi-emperical model based on the Lommel-Seeliger law. The lunar flux, 
normalized to unity at full moon, was found to be reasonably well-fit by the equation
\begin{eqnarray}
	f & = & \{1+\ln [\tan (\varphi /4)]\sin (\varphi /2)\tan (\varphi /2)\} \nonumber \\
	& & \times (0.4 e^{-0.09\varphi} + 0.6e^{-0.01\varphi}) \label{eq:flux}
\end{eqnarray}
where $\varphi$ is the lunar phase angle in degrees. After applying a correction for airmass, 
the observed flux was compared to that predicted by Eqn. (\ref{eq:flux}) and the 
ratio of observed to predicted flux was computed. The result gives an estimate of 
atmospheric transparency that is generally accurate to within about 10\%.

This transparency index can be used to estimate the fraction of clear sky at the site. 
For each run, the number of records for which the transparency is estimated to be 
greater than 50\% was determined, and divided by the number of records that were 
frost-free. The result is reported in column (5) of Table 1 as the fraction of ÒclearÓ sky.

The definition of photometric conditions is somewhat arbitrary, depending on the 
desired photometric accuracy. For this paper, we define photometric to mean 
that the observed scintillation index of the Moon does not exceed $10^{-5}$. This
is more than an order of magnitude larger than the scintillation index that would
result from atmospheric turbulence alone. It corresponds to RMS atmospheric
transparency variations at the level of 0.3\%, i.e. 0.003 magnitudes.

In practice this means that the sky is clear enough for us to obtain an accurate 
turbulence profile. However, thin cirrus clouds might still be present. Due to 
their high ($\sim 10$ km) altitude, such clouds produce intensity correlations 
over all baselines. The resulting covariance is practically independent of 
baseline over the range
covered by our instrument. In contrast, the atmospheric covariance signal, which
arises primarily from turbulence in the lower km of the atmosphere, falls rapidly with
increasing baseline and is essentially zero on the largest baselines.
Transparency fluctuations are independent of the atmospheric
turbulence signal, so the covariances of the two effects are additive. Therefore,
in the case of very thin cirrus ($< 0.3$\% optical depth), the atmospheric covariance 
function can be recovered by subtracting a constant background determined from
the longest baseline.

From all recorded data, covariances were computed for 120 s intervals.
For each 120 s sequence, the mean transparency, ice index and 
scintillation index was also computed. Sequences
that were deemed to be free of frost or ice, and were found to be photometric
by the above criteria were analyzed by our standard lunar scintillometer
pipeline. 

The inversion technique is described in \cite{h2008}. Briefly, an 8-parameter
model was used to represent the atmospheric turbulence profile. For each
set of model parameters, the predicted covariance for each baseline was
computed and a nonlinear optimization technique was used to find the model
parameters that best fit the data. A Bayesian approach was used which 
enforces the condition that the model be physically acceptable (no negative 
values of $C_N^2$) 
and employs prior knowledge of the typical variations of $C_N^2$ found at
astronomical sites.  The optimization was done by maximizing the posterior
probability by means of a Markov-chain Monte Carlo algorithm. 
The model parameters are the values of $C_N^2$ at eight reference 
heights, which are chosen to be 1, 10, 20, 40, 80, 160, 320, and 640 m.
Linear interpolation in log-log space was used to determine  $C_N^2$ values 
between the reference heights. Above 640 m, the model decreases 
as the -1.735 power of the height. As will be shown below, this function
provides a good fit to the observed data in the range 10 - 640 m. The exact 
behaviour at high altitudes is not critical since the scintillometer has little 
sensitivity at altitudes above $\sim 1$ km, due to a combination of aperture 
averaging and the effects of the outer scale of turbulence. As a check,
the pipeline was run with a rather extreme model which dropped rapidly
above 640 m, with a power law index of -9. This changed the derived
values of the seeing FWHM by less than 2\%.

The covariance on baseline $r$ (the component of separation perpendicular
to the line of sight) is related to $C_N^2(h)$ by an integral equation
\begin{equation}
	B(r) = \int_0^\infty C_N^2(z \cos \zeta) W(r,z) dz
\end{equation}
where $\zeta$ is the zenith angle of the Moon and $W(r,z)$ is a response 
function that represents the covariance on baseline $r$ produced by a unit 
impulse at range $z$. This function depends on the phase and orientation
of the Moon, the size and shape of the sensor active area and the assumed
turbulence outer scale $L_0$. Response functions were precomputed for 
all baselines, for every degree 
of lunar phase.  The outer scale was taken to be 20 m.

Once the $C_N^2$ profile has been determined, Fried's parameter
$r_0$ and the FWHM $\varepsilon$ of the atmospheric point-spread 
function  at the zenith can be computed, as a function 
of the height of the telescope above the ground, by integrating the $C_N^2$ 
profile vertically through the atmosphere according to 
\begin{eqnarray}
	r_0(h) & = & (16.70 \lambda^{-2})^{-3/5}\int_h^\infty C_N^2(z) dz \\
	\varepsilon(h) & = & 0.976 \lambda /r_0
\end{eqnarray}
As is standard, the results were computed for a wavelength of 
$\lambda = 0.5$ um and a zenith angle of $0^\circ$.

\section[]{Results}

A total of 3118 2-minute records, corresponding to 103.93 hours of observation, 
were photometric and frost-free. 
From these, 3118 $C_N^2$ profiles were determined. The median values
of $C_N^2$ are shown by the points in Fig. 3. The line 
shown in the figure is the best fit linear regression to the data points. It 
corresponds to the equation
\begin{equation}
	C_N^2(h) = 1.856\times 10^{-12}h^{-1.735}
\end{equation}
where $h$ is the height above ground in m and the units of $C_N^2$ are 
m$^{-2/3}$. It can be seen that this power law fits the data quite well for 
heights above 16 m. The median, as well as 10, 25, 75 and 90th 
percentiles of $C_N^2$ are plotted as a function of height, in the lower 
400 m of the atmosphere, in Fig. 4. During times of poor seeing, $C_N^2$ 
rises by about three orders of magnitude from 400 m to 10 m above
ground. In good seeing, the rise is about a factor of ten smaller.

\begin{figure}
\hspace{-3mm}
\includegraphics[width=94 mm]{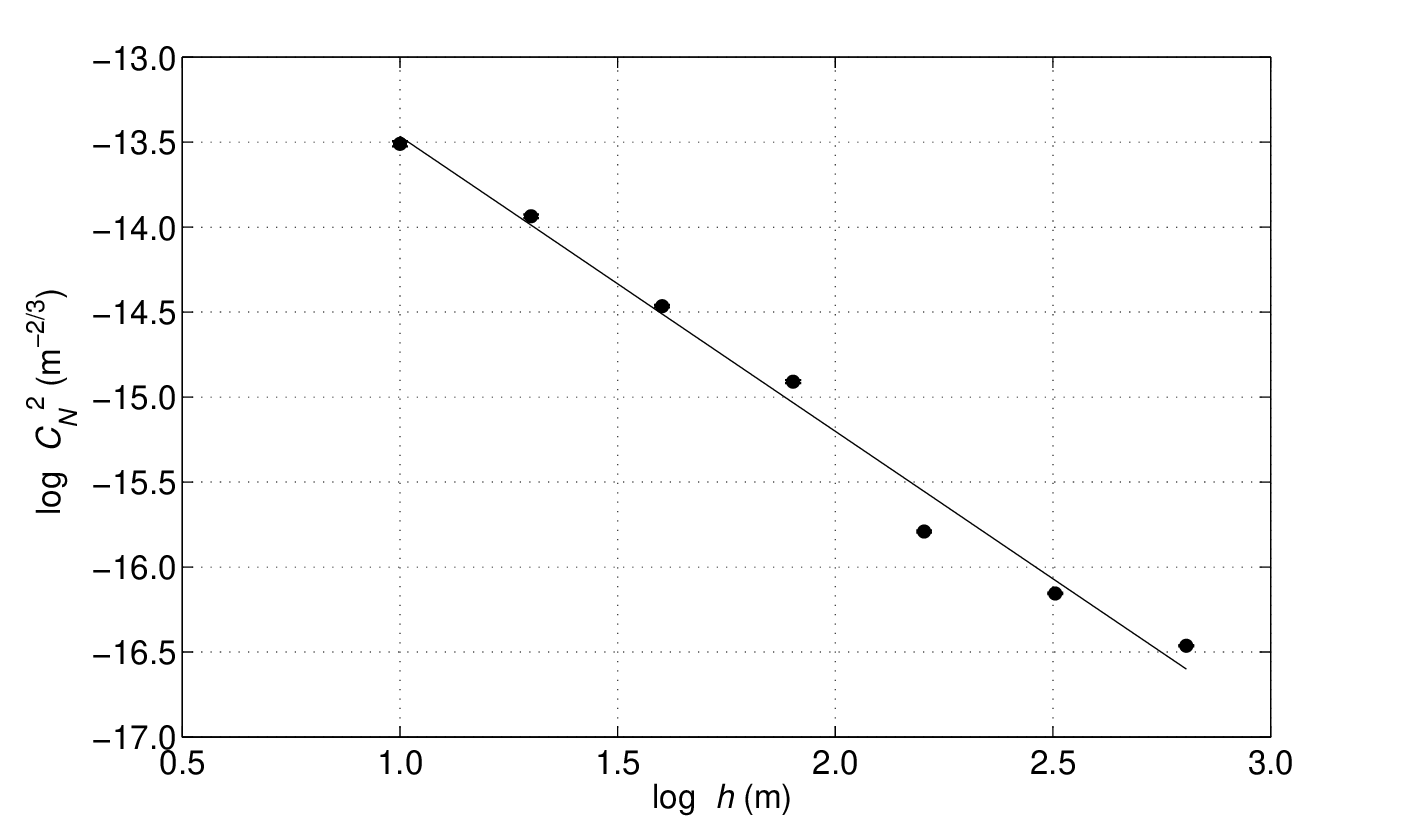}
%  \vspace{302pt}
 \caption{Median of 3118 measurements of $C_N^2$ at reference  heights 
ranging from 10 to 640m above the ATP instrument}
\end{figure}

\begin{figure}
\hspace{-2mm}
\includegraphics[width=92 mm]{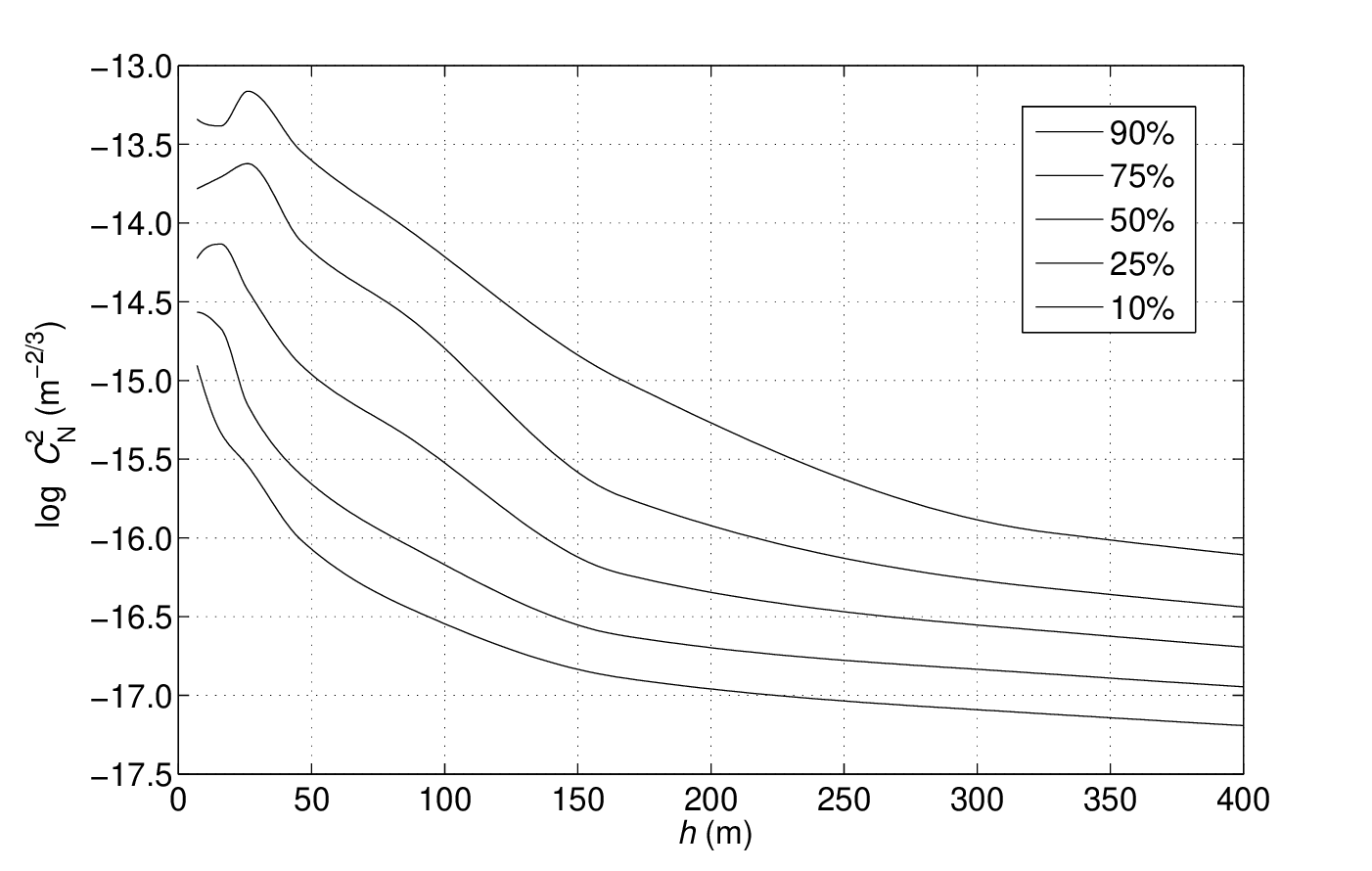}
%  \vspace{302pt}
 \caption{Percentiles of $C_N^2$ distribution as a function of height above
 ground, in the lower 400m of atmosphere. From top to bottom, the curves
 represent 90th, 75th, 50th, 25th and 10th percentiles.}
\end{figure}

A statistic of particular interest for astronomy is the full width at half-maximum 
intensity (FWHM), $\epsilon(h)$, of the atmospheric point spread function. 
This determines the image quality for seeing-limited observations with a 
telescope located at a height $h$ above ground. For large telescopes,
$h$ should be taken as the height of the dome, or dome shutters.
The cumulative distribution of $\varepsilon$, computed for heights 
ranging from 10 to 60 m above ground, is shown in Fig. 5. Percentiles of
these distributions are listed in Table 2, which also includes values for
a 7-metre altitude, to facilitate comparison with other astronomical sites.

In Fig. 6. the ground-layer seeing is shown as a continuous 
function of height above ground, for the same percentiles as in Fig. 4.
From this it can be seen that the ground layer is generally quite thin,
with the seeing at 40 m being better by about a factor of two compared
to 7 m.

\begin{figure}
\hspace{-6mm}
\includegraphics[width=97 mm]{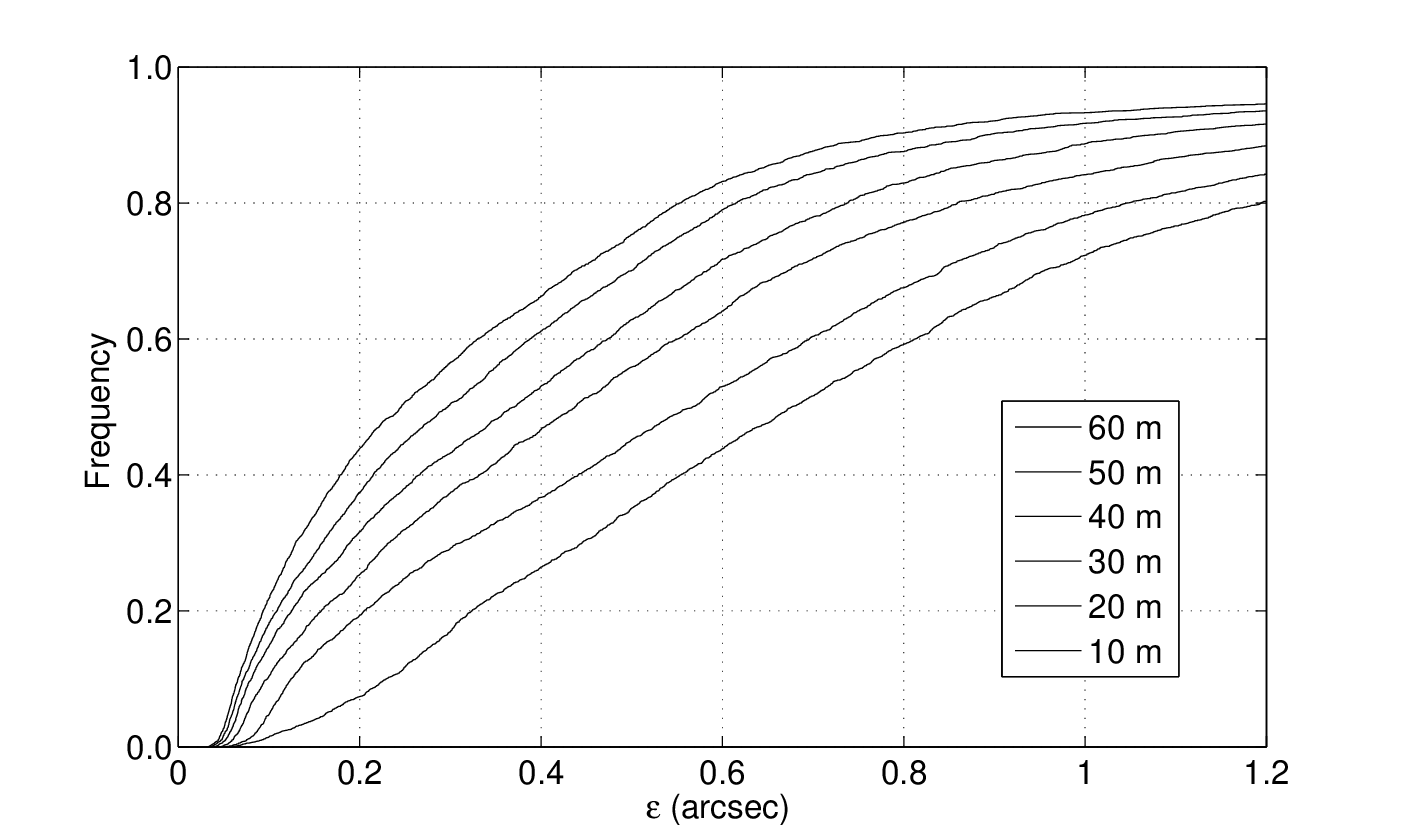}
%  \vspace{302pt}
 \caption{Cumulative distribution of predicted ground-layer seeing 
 FWHM for 6 different heights of the telescope above ground (from 
 top to bottom, 60 - 10 m).}
\end{figure}

\begin{figure}
\hspace{-6mm}
\includegraphics[width=97 mm]{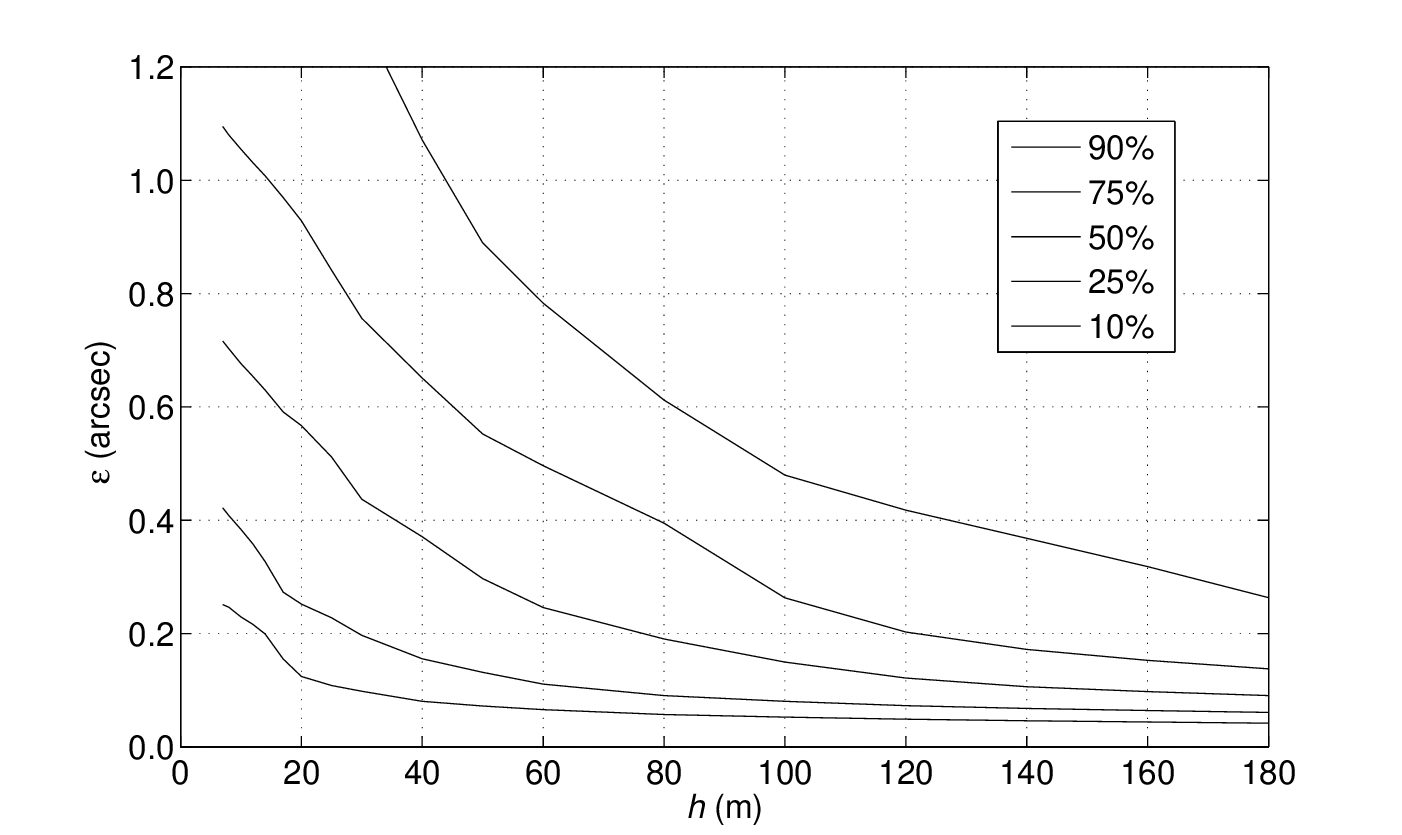}
%  \vspace{302pt}
 \caption{Percentiles of predicted seeing FWHM as a function of height
of the telescope above ground.}
\end{figure}

\begin{table}
 \centering
% \begin{minipage}{140mm}
  \caption{Percentiles of the distribution of ground-layer seeing 
  FWHM (arcsec) as seen from seven heights above ground.}
  \begin{tabular}{@{}lrrrrr@{}}
  \hline
  (1) & (2) & (3) & (4) & (5) & (6) \\
  Height (m) & 10\% & 25\% & 50\% & 75\% & 90\% \\
  \hline
    7 & 0.251 & 0.422 & 0.716 & 1.095 & 1.894 \\
  10 & 0.230 & 0.384 & 0.676 & 1.055 & 1.823 \\
  20 & 0.124 & 0.252 & 0.567 & 0.929 & 1.587 \\
  30 & 0.098 & 0.197 & 0.437 & 0.757 & 1.288 \\
  40 & 0.080 & 0.156 & 0.371 & 0.651 & 1.071 \\
  50 & 0.072 & 0.132 & 0.297 & 0.552 & 0.890 \\
  60 & 0.066 & 0.111 & 0.246 & 0.496 & 0.783 \\
\hline
\end{tabular}
% \end{minipage}
\end{table}

\section[]{Discussion}

The results described above apply only to the ground-layer component of atmospheric
turbulence, which comprises essentially all turbulence in the lower km of the atmosphere. 
In addition, there will be a contribution from the free atmosphere, which extends from
the ground layer to the stratosphere. The contribution of the free atmosphere to the seeing
FWHM depends primarily on turbulence in the tropopause region where wind shear 
associated with the jet stream creates strong turbulent layers. The total seeing can be
found by adding the turbulence integrals for the ground layer and the free atmosphere.
This leads to the ``quadrature'' relation
\begin{equation}
	\varepsilon = \left(\varepsilon_{GL}^{5/3}+\varepsilon_{FA}^{5/3}\right)^{3/5}
\end{equation}
for the ground layer and free atmosphere seeing, $\varepsilon_{GL}$ and $\varepsilon_{FA}$.

At the best mid-latitude 
sites, the median free-atmosphere seeing is on the order of 0.33 - 0.48 arcsec \citep{s2009}. 
In contrast, values closer to 0.25 arcsec have been found above the Antarctic plateau \citep{a2006}.
As our scintillometer is not sensitive to high-altitude turbulence, we are not able to measure
$\varepsilon_{FA}$. However, recent observations have been made using a Multi-Aperture 
Scintillation Sensor (MASS). From 118 hours of observations in the winter of 
2011-2012, \cite{s2013a} obtained typical values on the order of 0.30 arcsec for the free 
atmosphere above PEARL. 

For a given free atmosphere contribution, and the ground-layer seeing values from Table 2,
we estimate the total seeing using Eqn. (6). Taking $\varepsilon_{FA} = 0.30$ arcsec
from the MASS observations, we obtain the estimated total seeing shown in Table 3. 
These values are in good agreement with recent measurements made from the roof of
the PEARL building, using a differential image motion monitor \citep{s2013a}.

\begin{table}
 \centering
% \begin{minipage}{140mm}
  \caption{Percentiles of the distribution of estimated total seeing FWHM (arcsec) 
  as seen from seven heights above ground. A contribution of 0.30 arcsec due
  to free atmosphere turbulence has been assumed (Steinbring et al. 2013).}
  \begin{tabular}{@{}lrrrrr@{}}
  \hline
  (1) & (2) & (3) & (4) & (5) & (6) \\
  Height (m) & 10\% & 25\% & 50\% & 75\% & 90\% \\
  \hline
    7 & 0.419 & 0.552 & 0.813 & 1.169 & 1.946 \\
  10 & 0.404 & 0.521 & 0.776 & 1.131 & 1.877 \\
  20 & 0.340 & 0.419 & 0.678 & 1.011 & 1.646 \\
  30 & 0.327 & 0.382 & 0.565 & 0.850 & 1.355 \\
  40 & 0.319 & 0.357 & 0.510 & 0.753 & 1.146 \\
  50 & 0.316 & 0.344 & 0.452 & 0.664 & 0.975 \\
  60 & 0.314 & 0.333 & 0.415 & 0.615 & 0.874 \\
\hline
\end{tabular}
% \end{minipage}
\end{table}

Further insight can be obtained by examining records from the 
PEARL weather station for the times of our observations. These were
kindly provided to us by Dr. Pierre Fogal. 
In Fig. 1. we see that, during clear and frost-free hours of winter 
darkness, the wind rarely came from the west. The predominant
direction is from the SE. This puts the ATP instrument downwind
of the bulk of the heated laboratory building, so one might expect 
some thickening of the ground layer and degradation of seeing
as a result. Fig. 7 shows that the seeing is best when the 
wind is in the NE quadrant, and significantly worse in the SE 
quadrant. This pattern is also seen at the 50 m height, so is unlikely
to be due to a purely local effect. The seeing is somewhat worse when 
the wind is from the SW. This could be an effect of turbulence
induced by the ridge to the west of PEARL (Fig. 1), but we also
find relatively-good seeing for the azimuth range 
$300 - 315$ degrees. It would be unwise to draw any firm conclusions
for westerly winds due to the small quantity of data for the range
$180 - 330$ degrees. 

Wind speed can also affect seeing. A moderate wind is 
usually best, to carry away local thermal plumes while not inducing
too much mechanical turbulence. At PEARL, the wind speed 
correlates strongly with direction, at least for the times of our 
observations. Fig. 8 shows that the average wind speed is high, 
in the range $10-20$ m/s, when the wind direction is in the range
$100 - 160 $ degrees. This may also be a contributing factor to
the relatively-high seeing FWHM found in this same azimuth range.
There, seeing values exceeding 1 arcsec (at 7 m) correspond to
wind speed exceeding 15 m/s. However, wind speed does not
appear to be the cause of the high values of FWHM found when
the wind is in the western quadrants, since the wind speed from
those directions is relatively low, typically 5 m/s.

These results show that seeing at the PEARL site is often comparable 
to that found at mid-latitude sites that are at much higher altitudes. For example,
\cite{s2009} found median values of ground-layer seeing, from a
height of 7 m,  ranging from 0.34 to 0.58 arcsec  for the five candidate sites 
studied for the Thirty Meter Telescope project. The values of total 
seeing ranged from 0.63 to 0.79 arcsec at these sites.

\begin{figure}
\hspace{-7mm}
 \includegraphics[width=97 mm]{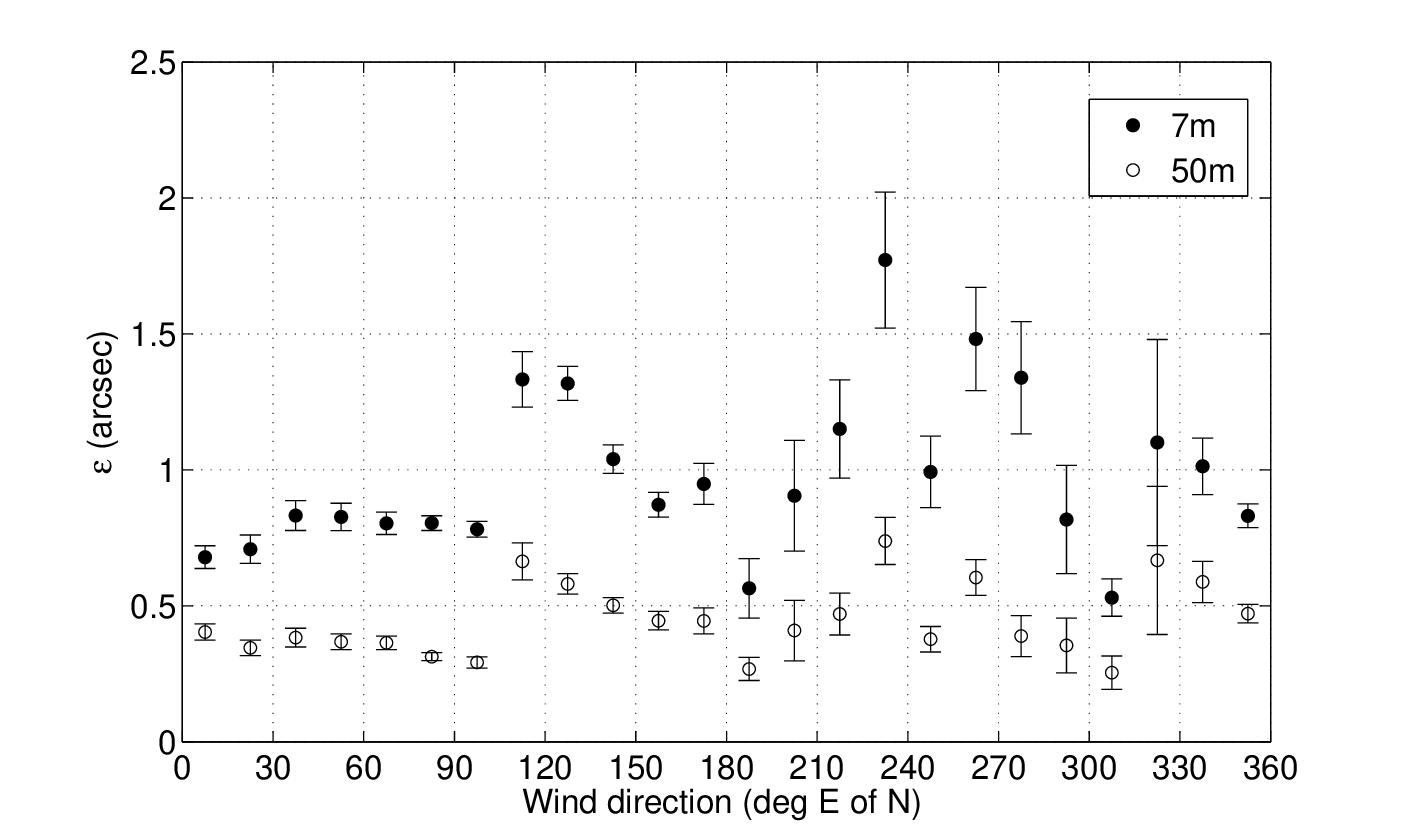}
%  \vspace{302pt}
 \caption{Mean ground-layer seeing FWHM at 7 and 50 m above ground as a function
 of wind direction, in 15-degree bins. The bars show standard errors.}
\end{figure}

\begin{figure}
\hspace{-7mm}
 \includegraphics[width=97 mm]{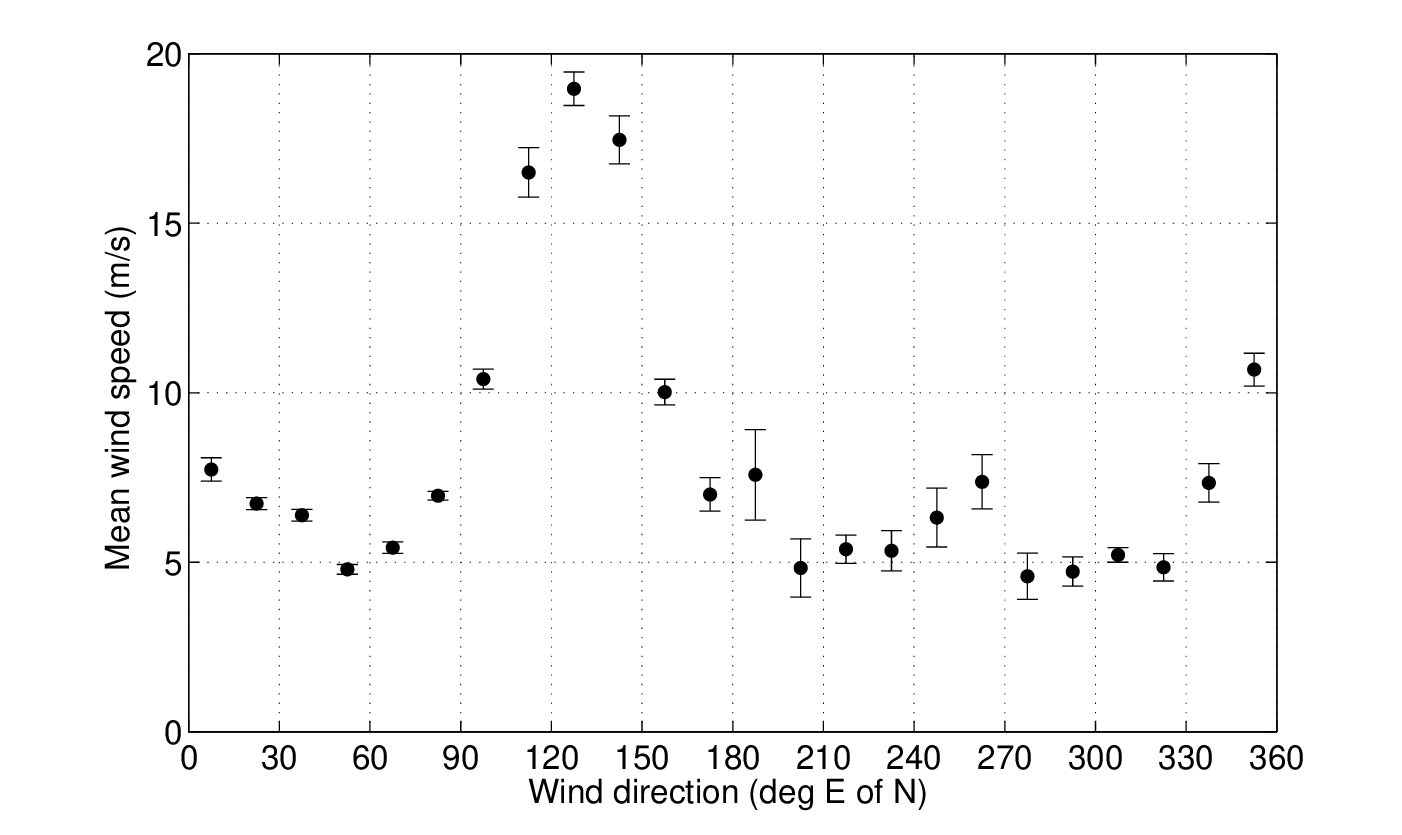}
%  \vspace{302pt}
 \caption{Mean mean wind velocity as a function of wind direction, during
the times of good weather when useful ATP data were obtained. }
\end{figure}

The seeing at PEARL is surprisingly good considering that its altitude is
only 610 m. There is higher terrain near Eureka and the mountain sites
along the north-western coast of Ellesmere Island ($82^\circ$N) 
investigated with weather stations and sky monitors have altitudes in the range
1400 to 1900 m.  These sites were all found to have climate conditions 
as good or better than PEARL. However, so far only some basic image 
quality measurements have been made with a Polaris tracker, not true 
seeing measurements \citep{s2013b}. Deploying
the ATP at one of these remote high-altitude sites would require a reliable 
source of electric power and improvements to the ice protection system.
But solutions to these technical issues are readily found, which could
allow measurements as part of an expanded exploration program.

\section*{Acknowledgments}

We thank J. Drummond and the staff of PEARL for providing access
to the facility, and for invaluable technical support. We are indebted to
P. Fogal for providing unpublished data from the PEARL meteorological
station for the dates of our observations. Logistical
support was provided by Environment Canada and by Natural Resources
Canada through the Polar Continental Shelf Program. TP and ES are
particularly indebted to the staff of the Eureka weather station for their
warm hospitality. M. Sch\"ock provided helpful advice and assistance 
with preparations at PEARL.  We thank the referee, T. Travouillon,
for his helpful comments. This work was supported by grants from 
the Natural Sciences and Engineering Research Council of Canada.

\end{document}